\documentclass[aps,prl,twocolumn,superscriptaddress]{revtex4-1}
\usepackage{amsmath}	%allows the use of mathematical symbols
\usepackage{graphicx}	% allows the use of the include graphics functions
\usepackage[usenames, dvipsnames]{color}
\usepackage{gensymb}
\usepackage{comment}
\normalfont

\usepackage{hyperref}
\usepackage{upgreek}
\newcommand{\micro}{${\upmu}$}

\begin{document}

\title{Valley addressable exciton-polaritons in atomically thin semiconductors}

\author{S. Dufferwiel}
\email{s.dufferwiel@sheffield.ac.uk}
\author{T. P. Lyons}
\affiliation{Department of Physics and Astronomy, University of Sheffield, Sheffield S3 7RH, UK}
\author{D. D. Solnyshkov}
\affiliation{Institut Pascal, Blaise Pascal University, 24 avenue des Landais, 63177 Aubiere, France}
\author{A. A. P. Trichet}
\affiliation{Department of Materials, University of Oxford, Parks Road, Oxford OX1 3PH, UK}
\author{F. Withers}
\affiliation{School of Physics and Astronomy, University of Manchester, Manchester M13 9PL, UK}
\affiliation{Centre for Graphene Science, CEMPS, University of Exeter, Exeter, EX4 4QF, UK}
\author{S. Schwarz}
\affiliation{Department of Physics and Astronomy, University of Sheffield, Sheffield S3 7RH, UK}
\author{G. Malpuech}
\affiliation{Institut Pascal, Blaise Pascal University, 24 avenue des Landais, 63177 Aubiere, France}
\author{J. M. Smith}
\affiliation{Department of Materials, University of Oxford, Parks Road, Oxford OX1 3PH, UK}
\author{K. S. Novoselov}
\affiliation{School of Physics and Astronomy, University of Manchester, Manchester M13 9PL, UK}
\author{M. S. Skolnick}
\author{D. N. Krizhanovskii}
\affiliation{Department of Physics and Astronomy, University of Sheffield, Sheffield S3 7RH, UK}
\author{A. I. Tartakovskii}
\email{a.tartakovskii@sheffield.ac.uk}
\affiliation{Department of Physics and Astronomy, University of Sheffield, Sheffield S3 7RH, UK}

\date{\today}

\begin{abstract}

% Valleytronic based abstract
While conventional semiconductor technology relies on the manipulation of electrical charge for the implementation of computational logic, additional degrees of freedom such as spin and valley offer alternative avenues for the encoding of information \cite{Yu2014}. In transition metal dichalcogenide (TMD) monolayers, where spin-valley locking is present \cite{Xiao2012,Mak2012}, strong retention of valley chirality has been reported for MoS$_2$, WSe$_2$ and WS$_2$ \cite{Sallen2012, Lagarde2014, Jones2013, SuzukiR2014,Zhu2014,Zhu2014_2} while MoSe$_2$ shows anomalously low valley polarisation retention \cite{Wang2015}. In this work, chiral selectivity of MoSe$_2$ cavity polaritons under helical excitation is reported with a polarisation degree that can be controlled by the exciton-cavity detuning. In contrast to the very low circular polarisation degrees seen in MoSe$_2$ exciton and trion resonances, we observe a significant enhancement of  up to 7 times when in the polaritonic regime. Here, polaritons introduce a fast decay mechanism which inhibits full valley pseudospin relaxation and thus allows for increased retention of injected polarisation in the emitted light. A dynamical model applicable to cavity-polaritons in any TMD semiconductor, reproduces the detuning dependence through the incorporation of the cavity-modified exciton relaxation, allowing an estimate of the spin relaxation time in MoSe$_2$ which is an order of magnitude faster than those reported in other TMDs. The valley addressable exciton-polaritons reported here offer robust valley polarised states demonstrating the prospect of valleytronic devices based upon TMDs embedded in photonic structures, with significant potential for valley-dependent nonlinear polariton-polariton interactions.

\end{abstract}


\maketitle

Single layers of TMDs are two-dimensional (2D) direct band-gap semiconductors which exhibit pronounced exciton resonances with binding energies of up to 0.55 eV \cite{MakPRL2010,Berkelbach2013,Ugeda2014}. The 2D nature of TMDs gives rise to strongly confined excitons with Bohr radii of around 1 nm, leading to a large oscillator strength of the excitonic transition with absorption as strong as 15\% \cite{MakPRL2010}. Recently, strong exciton-photon coupling has been reported in van der Waals crystals such as MoSe$_2$, WSe$_2$, WS$_2$ and MoS$_2$ embedded in optical microcavities \cite{Liu2015,Dufferwiel2015,Flatten2016,Sidler2016,Lundt2016_2}. The observation of exciton-polaritons in TMDs creates opportunities in engineering the polariton-polariton interaction \cite{Sturm2014,Felix2015,Walker2015} in 2D materials, which can potentially be used for the development of active nanophotonic devices through integration of thin films into photonic crystal cavities and waveguides \cite{Kim2012}. Moreover, TMD based polaritons additional degrees of freedom of valley pseudospin and finite Berry curvature can be utilised in new valleytronic implementations \cite{Yu2014}.

\begin{figure*}
\center
\includegraphics{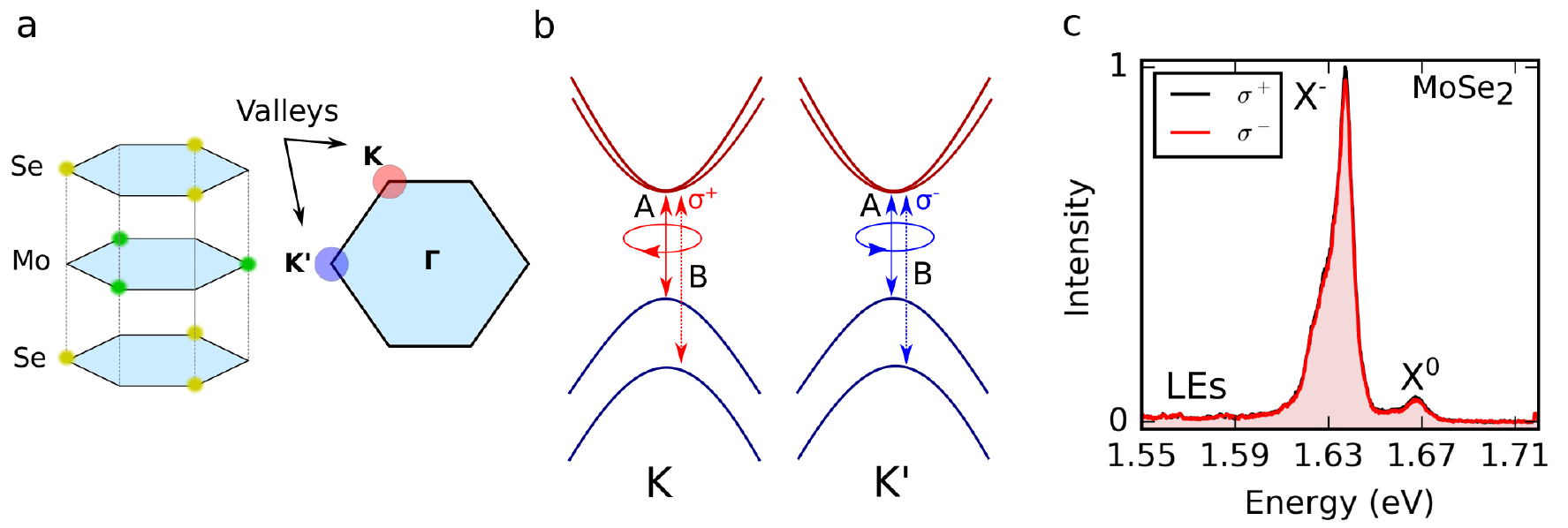}
\caption{\label{fig:figure1} \textbf{Spin-valley locking of excitons in MoSe$_2$.} \textbf{a} Illustration of the hexagonal lattice structure of MoSe$_2$. Optical band gaps are formed at the inequivalent K and K' valleys at the edges of the Brillouin zone. \textbf{b} Sketch of the valley dependent optical selection rules at the band edges. In the K (K') valley, the optical transitions couple to $\sigma^+$ ($\sigma^-$) polarised light. \textbf{c} PL spectra of MoSe$_2$ under $\sigma^+$ excitation showing A-exciton (X$^0$), trion (X$^-$) and localised emitter (LE) emission in co- (black) and cross- (red) polarised detection.}
\end{figure*}

In TMDs, the combination of time-reversal symmetry, strong spin-orbit coupling and the absence of crystal inversion symmetry leads to the coupling of spin and valley degrees of freedom \cite{Xiao2012,Mak2012}. At both the conduction and valence band edges, the electron spin orientation is locked to the valley index allowing direct optical initialization of the valley pseudo-spin in the non-equivalent K and K' valleys of the hexagonal Brilluoin zone with $\sigma^+$/$\sigma^-$ helicities respectively, as shown in Fig.~ \ref{fig:figure1}a and Fig. ~\ref{fig:figure1}b. Upon optical excitation the valley polarisation can be  directly probed through the degree of circular polarisation of the photoluminescence (PL) which is defined as $\rho = (I_{co}-I_{cross}) / (I_{co} + I_{cross})$ where $I_{co}$ ($I_{cross}$) is the component co-polarised (cross-polarised) relative to the excitation. Extensive polarisation-resolved PL studies of MoS$_2$, WSe$_2$ and WS$_2$ report high initialized degrees of circular polarisation, indicating strong chiral selectivity of excitonic states which are robust from inter-valley relaxation \cite{Sallen2012, Lagarde2014, Jones2013, SuzukiR2014,Zhu2014,Zhu2014_2}. Kerr rotation measurements on WSe$_2$ report exciton valley relaxation times of around 6 ps attributed to exciton intervalley transfer caused by precession in effective magnetic fields due to the long range electron-hole exchange interaction \cite{Zhu2014_2, Maialle1993, Glazov2014, Glazov2015, Yu2014_2}. The large valley polarisation degrees observed in continuous-wave (cw) PL studies are then possible since the exciton lifetime of around 2-5 ps also occurs on an ultrafast timescale \cite{Wang2014,Dufferwiel2015,Robert2016}. In stark contrast, A-excitons in MoSe$_2$, which have a comparable lifetime, show very low circular polarisation degrees of $<5$\% \cite{Wang2015}.  The exact mechanism leading to this anomalous polarisation behaviour of MoSe$_2$ is currently unclear, but would be consistent with  an unconfirmed sub-ps valley relaxation time.

In this article we report on the valley addressability of exciton-polaritons in MoSe$_2$ embedded in tunable microcavities.  A strong dependence of the polarisation degree on the exciton-cavity detuning is reported along with significant enhancement compared to the exciton and trion polarisation degrees. Due to interaction with disorder and localisation in the film, excitons undergo efficient ($< 0.1$ ps) scattering between low and high k-vectors where efficient depolarisation occurs due to the Maialle-Silva-Sham (MSS) mechanism - valley relaxation induced by scattering in the presence of exciton LT-splitting from the electron-hole exchange interaction \cite{Maialle1993, Glazov2014, Glazov2015,Yu2014_2}. This leads to extrinsic valley depolarisation and the observed low polarisation degrees of the bare exciton and trion emission in MoSe$_2$. Intrinsic depolarisation will also contribute, but the effect of disorder in current samples is dominant since the inhomogeneous broadening of 10-20 meV is much larger than the homogeneous linewidth. When in the strong exciton-photon coupling regime, the spatial extent of the polariton wavefunction ($1$ \micro m) is much greater than the typical length scale of exciton disorder (10s of nm), leading to significant reduction of polariton scattering by the exciton disorder potential \cite{Whittaker1996}. In addition, polaritons are detuned from the exciton reservoir states (lower and middle polariton branches) or have a photonic component (upper polariton branch) so there is a non-zero net relaxation of high k-vector exciton states to the polariton states. This relaxation to the polariton branches inhibits full valley pseudospin relaxation in the exciton reservoir resulting in an increased polarisation degree of the polariton emission. A dynamical model reproduces this dependence allowing estimates of the valley pseduospin relaxation time for bare excitons and trions to be $0.15$ ps and $1$ ps respectively. This sub-ps exciton intervalley relaxation is an order of magnitude faster than those reported for other TMDs providing strong evidence that the valley chirality in MoSe$_2$ is lost due to extremely fast pseudospin relaxation of excitons. The MSS depolarisation mechanism discussed here may be suppressed for bright excitons in TMDs with lowest energy dark exciton states, if the relaxation to dark states occurs on a comparable sub-ps timescale to disorder scattering. This provides an explanation for the discrepancy in polarisation behavior of MoSe$_2$ when compared with WSe$_2$ and WS$_2$ \cite{Zhu2011,Liu2013,K2013} as well as MoS$_2$, which was recently shown to have a lowest energy dark exciton state \cite{Echeverry2016}. The optically bright nature of the lowest energy exciton transitions in MoSe$_2$ combined with narrow linewidths make MoSe$_2$ the most suitable TMD for reaching the high polariton density regime and the achievement of nonlinear TMD polariton phenomena. The demonstration of retention and control of valley polarisation and suppression of intervalley relaxation for MoSe$_2$ polaritons opens new avenues for the exploration of spin/valley dependent polariton-polariton interactions in TMD microcavity systems as well as the development of valleytronic devices based upon TMDs embedded in photonic structures. 

The van der Waals (VDWs) heterostructure utilised in this study consists of a single monolayer of MoSe$_2$ encapsulated on both sides with thin films of hexagonal boron nitride (hBN) placed on the surface of a planar distributed Bragg reflector (DBR). PL spectra under non-resonant excitation at 1.946 eV at 4.2 K for MoSe$_2$ are shown in Fig.~ \ref{fig:figure1}c. Two characteristic peaks ascribed to neutral (X$^0$) and charged (X$^-$) exciton resonances are observed. In addition, localised emitters (LEs) are present on the low energy shoulder of the trion and below \cite{Chakraborty2016,Branny2016}. Under $\sigma^+$ excitation the polarisation resolved PL exhibits low retention of injected valley pseudospin of around 5\% and 2\% for the exciton and trion respectively, in agreement with previous reports \cite{Wang2015}. 

\begin{figure*}
\includegraphics{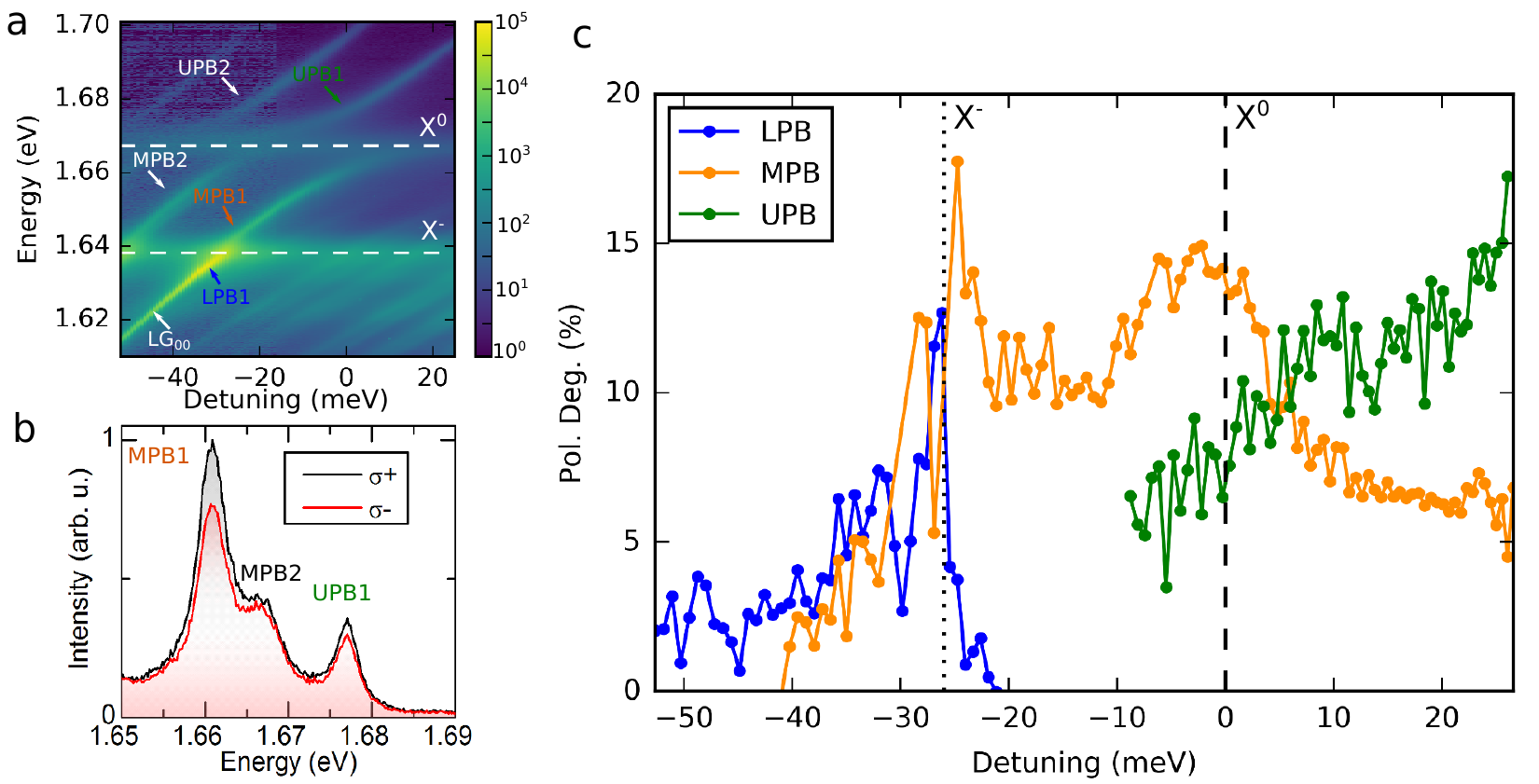}
\center
\caption{\label{fig:figure2} \textbf{Valley addressable exciton-polaritons in MoSe$_2$.} \textbf{a} Colourmap of PL spectra as a function exciton-photon detuning ($\Delta = E_c - E_{X^0}$), where $E_c$ and $E_{X^0}$ are the tunable cavity and exciton energies, showing clear anticrossing with X$^0$ with a Rabi splitting of $15.2 \pm 0.1$ meV. The onset of strong coupling is present for the trion with a coupling strengh of $1.3 \pm 0.1$ meV. \textbf{b} PL spectra at zero exciton-photon detuning with $\sigma^+$ excitation. Clear retention of circular polarisation is seen to varying degrees in all peaks. \textbf{c} Circular polarisation degree of the polariton branches as a function of exciton-photon detuning. The vertical dashed and dotted lines correspond to zero exciton-cavity and zero trion-cavity detuning.}
\end{figure*}

The tunable cavity is formed by introducing a top concave DBR into the optical path using nanopositioners and bringing the two DBRs to a total optical cavity length of around 2.9 \micro m \cite{Dolan2010,Trichet2015}. Full details of the cavity can be found in \cite{Dufferwiel2015}. The cylindrical symmetry of the formed hemispherical cavity introduces a harmonic-like photonic confinement leading to the formation of Laguerre-Gaussian (LG) cavity modes. A clear anticrossing is observed in PL when the modes are tuned through resonance with X$^0$ as shown for the longitudinal mode in Fig.~ \ref{fig:figure2}a. Coupling to the higher order modes is also present within the scan but all following discussion refers to the polariton branches formed due to coupling with the longitudinal mode. Fitting the polariton peak positions with a coupled oscillator model gives a Rabi splitting of  $15.2 \pm 0.1$ meV for X$^0$.  Analysis and fitting of the lineshape when close to resonance between the cavity mode and trion reveals the onset of strong coupling with a coupling strength of around $1.3 \pm 0.1$ meV as shown in Supplementary Fig. S1. The trion-polariton peaks are not fully resolved at resonance due to the cavity-trion Rabi splitting being comparable to the inhomogeneously broadened polariton linewidths.  In reflectivity and transmission, a clear splitting at resonance and the formation of trion-polariton branches has previously been shown \cite{Dufferwiel2015}. In the following discussion we take the peak below the trion as the lower polariton branch (LPB), the peak between the trion and exciton as the middle polariton branch (MPB) and the peak above the exciton as the upper polariton branch (UPB), as labeled in Fig.~ \ref{fig:figure2}a and customary for a strongly coupled 3-level system.

The retention of valley pseudospin of the polaritonic system was probed through excitation with non-resonant $\sigma^+$ circularly polarised light at 1.946 eV, close to a DBR stopband minimum. Fig.~ \ref{fig:figure2}b shows the PL spectra at zero exciton-photon detuning for co- (black) and cross-polarised (red) collection, where clear retention of valley polarisation for the MoSe$_2$ polariton branches occurs indicating an injected imbalance in the polariton valley populations. In-situ tunability of the cavity resonance allows the dependence of the polarisation degree on detuning to be probed through control of the mirror separation. The result for each polariton branch is plotted in Fig.~ \ref{fig:figure2}c  where the LG$_{00}$ cavity mode is tuned through resonance with X$^-$ and X$^0$ as in Fig.~ \ref{fig:figure2}a, and the polarisation degree is calculated from the peak intensities of fitted spectra in co- and cross-polarised components. Spectra corresponding to a full range of detunings are shown in Supplementary Fig. S2 and Fig. S3. For the LPB we see a low polarisation degree when strongly negatively detuned and an abrupt increase to around 15\% when close to resonance with X$^-$, before falling when positively detuned to the trion resonance. For detunings $\Delta > -20$ meV, coupling between the trion to high order modes masks the polarisation degree of the trion-like LPB, so is neglected in the following analysis. The MPB polarisation degree is initially low at negative detunings before rising to a maximum close to zero trion-cavity detuning, before dropping and increasing to a maximum of around $15$\% when slightly negatively detuned from X$^0$, before falling to the bare exciton level at positive detunings. Finally the UPB shows an almost linear increase in polarisation degree with detuning, to a maximum value of around 17\% at the maximum probed positive detuning of $\Delta = + 25$ meV. For all polariton branches, the maxima of polarisation degree are significantly enhanced in comparison to the bare exciton and trion resonances. Similar behaviour is measured for a second MoSe$_2$ sample with polarisation degrees of up to $20$\% as shown in Supplementary Fig. S5.

\begin{figure*}
\includegraphics{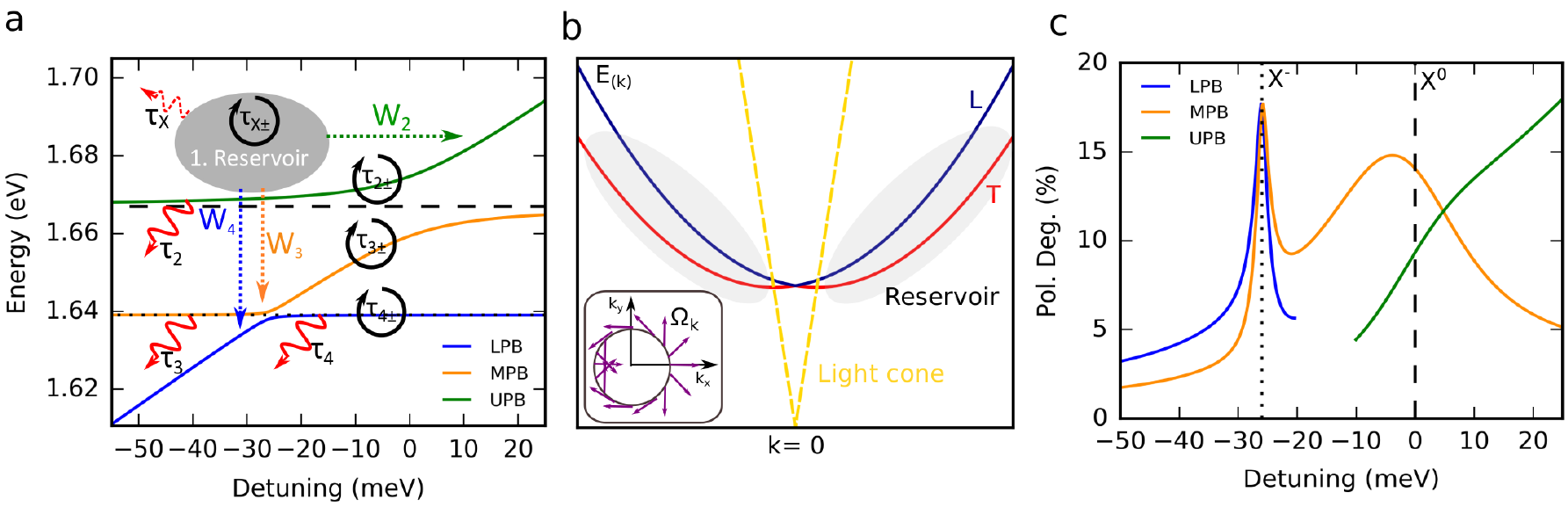}
\center
\caption{\label{fig:figure3} \textbf{Cavity modified relaxation dynamics.} \textbf{a} Scheme of the relevant states and relaxation dynamics.  \textbf{b} Illustration of the spin-splitting of the exciton dispersion into linearly polarised states with dipole moment parallel (L) and perpendicular (T) to the in-plane wavevector. Yellow dashed lines mark the light cone. Inset shows the orientation of the effective magnetic field $\Omega_k$ which induces spin precession in the exciton reservoir. \textbf{c} Calculated circular polarisation degree of the three polariton branches as a function of detuning using exciton and trion spin relaxation times of $\tau_{X\pm} = 0.15$ ps and $\tau_{t\pm} = 1$ ps respectively. The vertical dashed and dotted lines correspond to zero exciton-cavity and zero exciton-cavity detuning.}
\end{figure*}

To describe the processes taking place in the system with both exciton and trion resonances coupled with a single cavity mode we consider the dynamical process represented in Fig.~ \ref{fig:figure3}a. Excited valley polarised carriers relax to the reservoir of high k-vector excitons. The exciton depolarisation occurs through the MSS mechanism \cite{Maialle1993, Glazov2014,Glazov2015,Yu2014_2}. The exciton valley pseudospin precesses about the k-dependent effective magnetic field shown in the inset of Fig~ \ref{fig:figure3}b, which is induced by the exciton exchange coupling. This precession combined with random disorder scattering leads to exciton valley pseudospin relaxation with a characteristic time $\tau_{X\pm}$. Reservoir excitons, with non-radiative lifetime $\tau_{X}$ then relax towards the light-cone and form polaritons with a relaxation rate $W_i$. Further valley pseudospin relaxation can then occur in the polariton branches with a timescale $\tau_{i\pm}$ before decay out of the cavity with a polariton radiative lifetime $\tau_i$.  The resulting rate equations can be found in the Theoretical Model section below. The exciton-photon detuning affects the circular polarisation degree of the polariton emission via the excitonic ($x_i$) and trionic ($t_i$) fractions of the polariton states determining the scattering rates towards these states, while the lifetimes of the states are affected via their photonic fraction ($p_i$). The following dependences for the scattering rates to the middle and lower polariton branches (i = 3, 4), derived in Supplementary Note 1, are given by:

\begin{equation}
W_i = W_0(x_i+t_i)p_i
\label{eq:relaxation_rate}
\end{equation}

\noindent
where $W_0$ is the exciton energy relaxation rate. The scattering rates have to be proportional to the exciton fraction, and the exciton and trion are summed over, where we assume the presence of an extra electron does not change the relaxation mechanism. If the branch is completely excitonic it becomes degenerate with the exciton reservoir, and the scattering due to interaction with disorder towards the branch becomes balanced by an inverse scattering rate from the branch. The imbalance between the two scattering rates (forward and backward) $W_{X\to i}-W_{i\to X}\propto 1-e^{-(E_X-E_i)/k_B T}\approx (E_X-E_i)/k_B T$ is therefore proportional to the photonic fraction in the first order approximation, hence $W_i \propto p_i$.

The UPB is a special case, as it is always resonant with some states of the exciton reservoir at high k-vectors. It is therefore possible for high momentum excitons to scatter directly to the photon-like UPB states due to interaction with disorder. The corresponding scattering rate is therefore only proportional to the photonic fraction, while the excitonic one does not play any role: i.e. the exchange between the exciton reservoir and the excitonic fraction of this branch is always strongly balanced. Moreover, since the excitons do not need to relax their energy, and need only to change their momentum, the characteristic time of this process is that of momentum relaxation (pseudospin relaxation), and not that of energy relaxation. Therefore, we assume that:

\begin{equation}
W_2 \propto p_i/\tau_{X\pm}
\end{equation}

\noindent

Another consideration for the UPB is that the direct scattering from the reservoir not only drains it, shortening its lifetime and increasing the average polarisation degree, but also brings strongly polarised particles from higher energy states directly into the resonant polariton states. Since the spin precession rate is linearly proportional to the k-vector, we assume that the polarisation degree in the reservoir decreases linearly with energy, from 100\% at the injection point ($E_X+300$ meV) down to $\rho_X$ at $E_X$. This polarisation degree can then decay in the UPB before being emitted out of the system, giving a renormalization  $\tau_{2\pm}/\left(\tau_{2\pm}+\tau_2\right)$, derived in Supplementary Note 2. In order to reproduce the data the ratio of UPB polariton population coming from this direct scattering from the reservoir compared to indirect relaxation (Eq.~\ref{eq:relaxation_rate}) is required to be $1:1$.

We take the exciton and trion lifetimes as $\tau_{X0} = 5.3$ ps and $\tau_T = 12$ ps  \cite{Dufferwiel2015} and assume that the non-radiative exciton reservoir lifetime is comparable to the exciton PL-lifetime ($\tau_X = 5.3$ ps). The photon lifetime measured from the cavity Q-factor at large negative detuning is $\tau_p = 1.3$ ps. Finally we take the exciton energy relaxation rate to be $W_0 = 1/0.15$ ps to reproduce the data; an upper limit of $1$ ps can be justified from the resolution-limited rise time for MoSe$_2$ excitons in recent time resolved measurements \cite{Robert2016}. The polarisation dependence on detuning, shown in Fig.~ \ref{fig:figure3}c, is well reproduced using exciton and trion valley relaxation times of $\tau_{X\pm} = 0.15$ ps and $\tau_{T\pm} = 1$ ps respectively.  

The enhancement in polarisation degree of the lower and middle polariton branches relative to the bare monolayer case is due to cavity modified relaxation which allows MoSe$_2$ excitons to relax from the reservoir into polariton states quickly, inhibiting complete spin relaxation in the reservoir. In these polariton states, which are formed with excitons around $k \approx 0$, the exciton LT-splitting is small and hence the spin relaxation from the excitonic component of the polariton is around $20$ times slower than the bare exciton, as explained below in the Theoretical Model section. This is much slower than the polariton radiative lifetime, allowing lower and middle polariton branch particles to decay radiatively without further significant depolarisation, and with higher retention of valley pseudospin. When the photonic fraction of the polariton is high, relaxation into these polariton branches is less efficient causing larger accumulation of particles in the reservoir and hence low polarisation degree. At high excitonic fractions, scattering from the reservoir into these polariton branches is efficient, but the low photonic fraction, which determines the radiative decay and hence the imbalance between forward and backward scattering between polariton and reservoir states, leads to a longer time spent in the reservoir and hence low polarisation of emitted light. At intermediate detunings and roughly equal exciton and photon fractions a maximum in polarisation degree is seen which is a comprimise of efficient relaxation due to the exciton component and efficient radiative decay due to the photonic component. The high degree of valley polarisation in the UPB is due to direct scattering of high momentum excitons through interaction with disorder, populating the UPB with highly polarised particles which decay quickly due to their photonic component.

In conclusion, we observe clear retention of valley polarisation in MoSe$_2$  exciton-polaritons. The strong dependence on exciton-trion-photon components can be well described by a dynamical model that incorporates modified relaxation and state lifetimes arising from strong coupling with the cavity mode. We estimate that the valley pseudospin relaxation time for MoSe$_2$ excitons is sub-ps, indicating that the low polarisation retention of MoSe$_2$ is due to fast precession of exciton pseudospin caused by the MSS mechanism. This depolarisation  caused by disorder scattering is likely suppressed in TMDs with low energy dark states due to the additional fast decay channel, providing insight into the anomolous behavior of MoSe$_2$ in comparison to WSe$_2$, WS$_2$ and MoS$_2$. The spin-valley locking inherited by TMD polaritons from their excitonic component offers new avenues for valley-dependent polariton-polariton interactions \cite{Vladimirova2010} where the populations of polaritons in K and K' valleys can be controlled with circularly polarised excitation. This allows the utlisation of the valley degree of freedom in polariton condensates \cite{Kasprzak2006}, the optical spin hall effect \cite{Kavokin2005}, optical spin switching \cite{Amo2010} and polarisation bistabilities \cite{Sarkar2010}.

\section{Methods}

\subsection{Theoretical Model}

The Boltzmann equations for the populations of the various states are given as:

\begin{equation}
\frac{dn_{X_{\pm}}}{dt} = P_+ - \frac{n_{X_\pm}}{\tau_X} - n_{X_\pm} \sum\limits_{i=2}^4 W_i \mp \frac{n_{X_+} - n_{X_-}}{\tau_{X_\pm}}
\end{equation}

\begin{equation}
\frac{dn_{i_\pm}}{dt} = W_i n_{X_\pm} - \frac{n_{i_\pm}}{\tau_i} \mp \frac{n_{i_+} - n_{i_-}}{\tau_{i_\pm}}, i = 2,3,4
\end{equation}

\noindent
Here $n_{X\pm}$ and $n_{i\pm}$ are the exciton and polariton populations in the K and K' valleys respectively, where $i$ runs from 2 to 4 and denotes the UPB, MPB and LPB. $\tau_{i\pm}$ is the spin relaxation time of the polariton branch given below, and $P_+$ is the pumped valley-polarised population of particles. Solving these coupled rate equations in the steady state gives the solutions for the circular polarisation degrees of the exciton-polariton branches:

\begin{equation}
\rho_i = \frac{\tau_{i\pm}\tau_{X\pm} \Bigg(1 + \tau_{X} \sum\limits_{j=2}^4 W_j\Bigg)}{(2\tau_i + \tau_{i\pm}) \Bigg(\tau_{X\pm} + 2\tau_{X} + \tau_{X} \tau_{X\pm} \sum\limits_{j=2}^4 W_j \Bigg)}
\end{equation}

\noindent
For the exciton reservoir, the circular polarisation degree is given by:

\begin{equation}
\rho_X = \frac{\tau_{X\pm} \Bigg( 1 + \tau_{X} \sum\limits_{i=2}^4 W_i \Bigg)}{\tau_{X\pm} + 2\tau_{X} + \tau_{X} \tau_{X\pm} \sum\limits_{i=2}^4 W_i}
\end{equation}

The circular polarisation degree for the polariton branches and for the system in general is given by the balance of the exciton spin relaxation time and exciton energy relaxation time (total scattering rate out of the exciton reservoir). If the average time the particle spends in the reservoir is longer than the spin relaxation time, the circular polarisation degree will be low. 

The coupling of the exciton, trion, and photon modes is described by a coupled oscillator model, where the eigenenergies and the Hopfield coefficients are obtained from the matrix:

\begin{equation}
	\begin{pmatrix}
	E_X & 0 & V_X\\
	0 & E_T  & V_T\\
	V_X & V_T & E_C \\
	\end{pmatrix}
	\label{eq:hamiltonian}
\end{equation}
\noindent
where $E_X$ = 1.667 eV and $E_T=1.639$ eV  are the exciton and the trion energies and $E_C$ is the photon energy, which varies from 1.720 to 1.580 eV. The Rabi splitting for the exciton is taken as $2V_X = 15.2$ meV, while for the trion we take 2$V_T = 1.3$ meV. Diagonalizing this matrix, one obtains the three polariton branches as shown in Fig. ~\ref{fig:figure2}a. The corresponding eigenvectors give the exciton, trion, and photon fractions $x_i$, $t_i$, $p_i$. The polariton radiative lifetime can be written as:

\begin{equation}
\frac{1}{\tau_i} = \frac{x_i}{\tau_{X0}} + \frac{t_i}{\tau_T} + \frac{p_i}{\tau_p} 
\end{equation}

\noindent
where $\tau_p$, $\tau_{X0}$ and $\tau_{T}$ are the photon, exciton and trion lifetimes respectively. Finally, for the polariton spin relaxation times we take the following dependence:

\begin{equation}
\frac{1}{\tau_{i\pm}} = \alpha \frac{x_i}{\tau_{X\pm}} + \beta \frac{t_i}{\tau_{T\pm}} + \frac{p_i}{\tau_{p\pm}}
\end{equation}

\noindent
where we assume different valley pseudospin relaxation times for excitons, $\tau_{X\pm}$, and trions, $\tau_{T\pm}$. The coefficients $\alpha$ and $\beta$ are required since the polariton spin relaxation due to the excitonic component is reduced in comparison to the bare flake exciton and trion. In the bare flake efficient scattering between low k (small LT-splitting) and high k (large LT-splitting) excitons occurs due to scattering on disorder or localisation giving rise to fast depolarisation. Since polaritons are formed from low-k excitons and scattering due to disorder is reduced for the polaritons we expect the contribution to the polariton depolarisation from the exciton component to be less than the bare exciton spin relaxation time. Assuming a thermal population of excitons in the reservoir and a polariton wavefunction $\approx 1$ \micro m \cite{Dufferwiel2014} and a linear dependence of the effective magnetic field with in-plane wavevector \cite{Glazov2014,Yu2014_2} we estimate $\alpha \approx \beta \approx 0.05$. We also take into account spin relaxation of the photonic component, $\tau_{p\pm} = 15$ ps due to the finite TE-TM splitting of the ground state mode \cite{Kavokin2005} (see Supplementary Fig. S6).

\subsection{Sample Preparation}

The hBN-MoSe$_2$ stacks on DBR substrates were fabricated by mechanically exfoliated bulk MoSe$_2$ crystal onto a polymer double layer commonly used for dry transfer methods  \cite{Kretinin2014}. The MoSe$_2$ single layer flake was then used to lift, by van der Waals forces  \cite{Kretinin2014}, a thin mechanically exfoliated hBN flake from a separate Si-SiO$_2$ substrate. The whole stack was then dropped down on the DBR substrate. The PMMA membrane along with the heterostructure stack was heated to 130 degrees to soften the PMMA followed by its removal in acetone then isoproponal. To preserve the MoSe$_2$ flake from environmental effects a second thin hBN flake was used to fully encapsulate the TMD.  All bulk crystals including hBN were acquired from HQGraphene.

\subsection{Optical Measurements}
Optical measurements were performed with samples held in a helium bath cryostat system at a temperature of 4.2K. Top and bottom DBRs were attached to XYZ nanopositioners with additional goniometer stages allowing tilt control of the bottom DBR. Optical excitation of the bare monolayer was possible by removing the top DBR from the optical path. All \micro-PL experiments were performed with continuous-wave (cw) excitation using a 638 nm laser diode, focused onto the sample with an achromatic lens. polarisation resolved measurements were performed using a combination of linear polarizer and a quarter waveplate in the excitation path, and quarter wave-plate, half wave-plate and linear polarizer in the collection path. PL  was collected by focusing onto a single mode fibre which was guided into a 0.75m spectrometer and a high sensitivity charged couple device.

\subsection{Microcavity}

The tunable microcavity with embedded TMD monolayer is formed using an external concave mirror to produce a zero-dimensional tunable cavity \cite{Dufferwiel2014}. The formed cavity electric-field profile is shown in Supplementary Fig. S7 with the monolayer placed at an electric-field antinode, and nanopositioners are used to control the cavity spectral resonance energy. The nominal radius of curvature of the concave mirror is 20 \micro m leading to a beam waist on the planar mirror of around 1 \micro m \cite{Dufferwiel2014}.

\section{Acknowledgements}

We thank the financial support of the Graphene Flagship and the EPSRC grants EP/M012727/1 and EP/J007544/1. A. A. P. T., D. N. K and J. M. S. acknowledge support from the Leverhulme Trust. F. W. acknowledges support from the Royal Academy of Engineering and K. S. N. from the Royal Society, EPSRC, US Army Research Office and ERC Grant Hetero2D.

\section{Author Contributions}

S.D and T. P. L carried out optical investigations. A. A. P. T. designed and fabricated the concave mirrors. F. W. fabricated the MoSe$_2$ samples. D. D. S. and G. M. carried out theoretical analysis. S. D. analysed the data and prepared the manuscript with contributions from all co-authors. J. M. S., K. N., M. S. S., D. N. K., and A. I. T. provided management of various aspects of the project. D. N. K proposed the idea of using an open-access microcavity system, developed by his group, for polariton studies with TMD monolayers.  A. I. T. conceived and oversaw the project.

\newpage

\widetext
\clearpage

\textbf{\large Supplementary Information for `Valley addressable exciton-polaritons in atomically thin semiconductors'}

%%%%%%%%%% Merge with supplemental materials %%%%%%%%%%
%%%%%%%%%% Prefix a "S" to all equations, figures, tables and reset the counter %%%%%%%%%%
\renewcommand{\thesection}{S\arabic{section}}
\setcounter{section}{0}
\renewcommand{\thefigure}{S\arabic{figure}}
\setcounter{figure}{0}
\renewcommand{\theequation}{S\arabic{equation}}
\setcounter{equation}{0}
\renewcommand{\thetable}{S\arabic{table}}
\setcounter{table}{0}
\renewcommand{\citenumfont}[1]{S#1}
\makeatletter
\renewcommand{\@biblabel}[1]{S#1.}
\makeatother

\onecolumngrid

\section{Supplementary Note 1: Derivation of the scattering rates}
In this section, we provide more details on the derivation of the equations of the main text.

First, we would like to describe the scattering between the exciton reservoir (index $X$) and a polariton state (index $2$). In general, the rate equations for the exchange between such states can be written as follows:

\begin{equation}
\begin{array}{l}
\frac{{d{n_X}}}{{dt}} =  - {W_{X \to 2}}{n_X} + {W_{2 \to X}}{n_2}\\
\frac{{d{n_2}}}{{dt}} =  + {W_{X \to 2}}{n_X} - {W_{2 \to X}}{n_2}
\end{array}
\end{equation}
where we have omitted all other terms, e.g. lifetime, spin relaxation, etc. The scattering mechanism can be the exciton-phonon interaction or the exciton-exciton interaction: in presence of a thermalized exciton reservoir, both give the same dependence on the temperature and energy difference:
\begin{equation}
\begin{array}{l}
{W_{X \to 2}} = {W_0}x\\
{W_{2 \to X}} = {W_0}x{e^{ - \frac{{{E_X} - {E_2}}}{{{k_B}T}}}}
\end{array}
\end{equation}
The excitonic fraction $x$ of the polariton state appears in the scattering rate because only the exciton can efficiently interact with phonons, or the other excitons.  The coefficient $W_0$ is the same for both scattering rates, but since in one case (scattering down) the energy (the phonon) is emitted and in the other (scattering up) -- absorbed, the corresponding probability factors are different. Our goal is now to simplify these expressions, in order to obtain an analytical solution. Indeed, if the polariton branch is far below the reservoir, the scattering backwards from this branch to the reservoir is small and can be safely neglected. However, if this branch is close to the reservoir, this scattering is not negligible, and we need to introduce a correction to the scattering rate $W_{X\to 2}$ in order to take this into account. Indeed, in this case the scattering rates $W_{X\to 2}$ and $W_{2\to X}$ become comparable, as well as the populations, and so it is possible to write simply the difference of both terms:
\begin{equation}
{W_{X \to 2}} - {W_{2 \to X}} = {W_0}x\left( {1 - {e^{ - \frac{{{E_X} - {E_2}}}{{{k_B}T}}}}} \right)
\end{equation}
which in the first order can be written as
\begin{equation}
{W_{X \to 2}} - {W_{2 \to X}} \approx {W_0}x\frac{{{E_X} - {E_2}}}{{{k_B}T}}
\end{equation}

On the other hand, at strongly positive detunings $\Delta\gg V$ ($V$ is one half of the Rabi splitting), both the photonic fraction $p_2$ and the relative energy of the polariton state with respect to the exciton $(E_X-E_2)/\Delta$ scale as $V^2/\Delta^2$, which allows to simplify the expression for the scattering rate by using the photonic fraction instead of the energy difference:
\begin{equation}
W_{X \to 2}^{eff} \approx {W_0}xp
\end{equation}
and the rate equation now contains a single term with a single effective scattering rate:
\begin{equation}
\begin{array}{l}
\frac{{d{n_X}}}{{dt}} =  - {W^{eff}_{X \to 2}}{n_X} \\
\frac{{d{n_2}}}{{dt}} =  + {W^{eff}_{X \to 2}}{n_X}
\end{array}
\end{equation}
In the main text, we omit the index $eff$ on the scattering rate.

The advantage of all these simplifications is that the final model with a reduced number of terms has a clear analytical solution for the  polarization degree.

\clearpage

\section{Supplementary Note 2: Special case of the UPB}
In the main text, we have discussed that one can assume an energy dependence of the  polarization degree within the reservoir: fully polarized particles are injected at high energy and they progressively lose their polarization as they relax in energy. Therefore, if the UPB is resonant with some particular energy state within the reservoir, there will be an incoming scattering rate towards this UPB with a certain polarization degree (higher than at the bottom of the exciton reservoir). These polarized polaritons of the UPB will then lose their polarization within this branch, before being emitted out of the system. To obtain their final polarization degree, let us consider the UPB as a separate system with a pumping $P$ (which describes the scattering from the reservoir) carrying a polarization degree $\rho$. The lifetime of the states of the UPB is $\tau_2$ and the spin relaxation time $\tau_{2\pm}$. The rate equations for this reduced system can be written as follows:

\begin{equation}
\begin{array}{l}
\frac{{d{n_ + }}}{{dt}} = \frac{{P\left( {1 + \rho } \right)}}{2} - \frac{{{n_ + }}}{\tau } - \frac{{{n_ + } - {n_ - }}}{{2{\tau _ \pm }}}\\
\frac{{d{n_ - }}}{{dt}} = \frac{{P\left( {1 - \rho } \right)}}{2} - \frac{{{n_ - }}}{\tau } + \frac{{{n_ + } - {n_ - }}}{{2{\tau _ \pm }}}
\end{array}
\end{equation}

The stationary solution of this system of equations is:
\begin{equation}
\begin{array}{l}
{n_ + } = \frac{{\tau \left( {P\tau  + P{\tau _ \pm } + P\rho {\tau _ \pm }} \right)}}{{2\left( {\tau  + {\tau _ \pm }} \right)}}\\
{n_ + } = \frac{{\tau \left( {P\tau  + P{\tau _ \pm } - P\rho {\tau _ \pm }} \right)}}{{2\left( {\tau  + {\tau _ \pm }} \right)}}
\end{array}
\end{equation}
which gives a simple expression for the polarization degree of the UPB:

\begin{equation}
\rho ' = \frac{{\rho {\tau _ \pm }}}{{\tau  + {\tau _ \pm }}}
\end{equation}
used in the main text.

\clearpage

\begin{figure}
\includegraphics{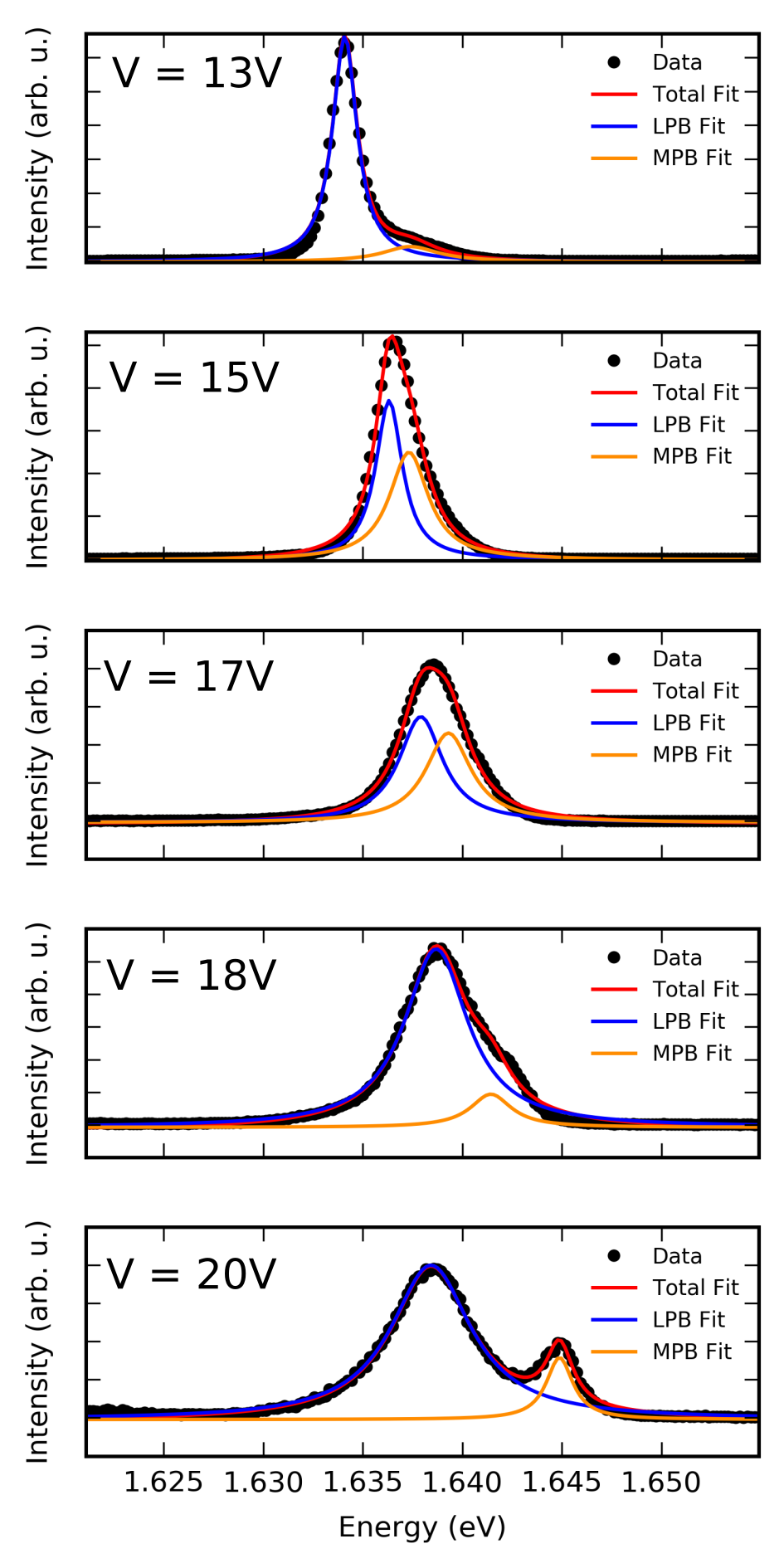}
\caption{\label{fig:trion_coupling} Spectra as a function of piezo voltage when the cavity is tuned through resonance with the trion emission energy. The presence of two peaks when slightly detuned and a strong asymmetry of the peak at resonance indicates the onset of strong coupling. Here the Rabi splitting of $1.3 \pm 0.1$ meV is comparable to the inhomogeneously broadened polariton linewidths preventing the polariton branches from being fully resolved at resonance. The fits correspond to those used to extract the polarization degree of the constituent polariton peaks.}
\end{figure}

\clearpage

\begin{figure}
\includegraphics{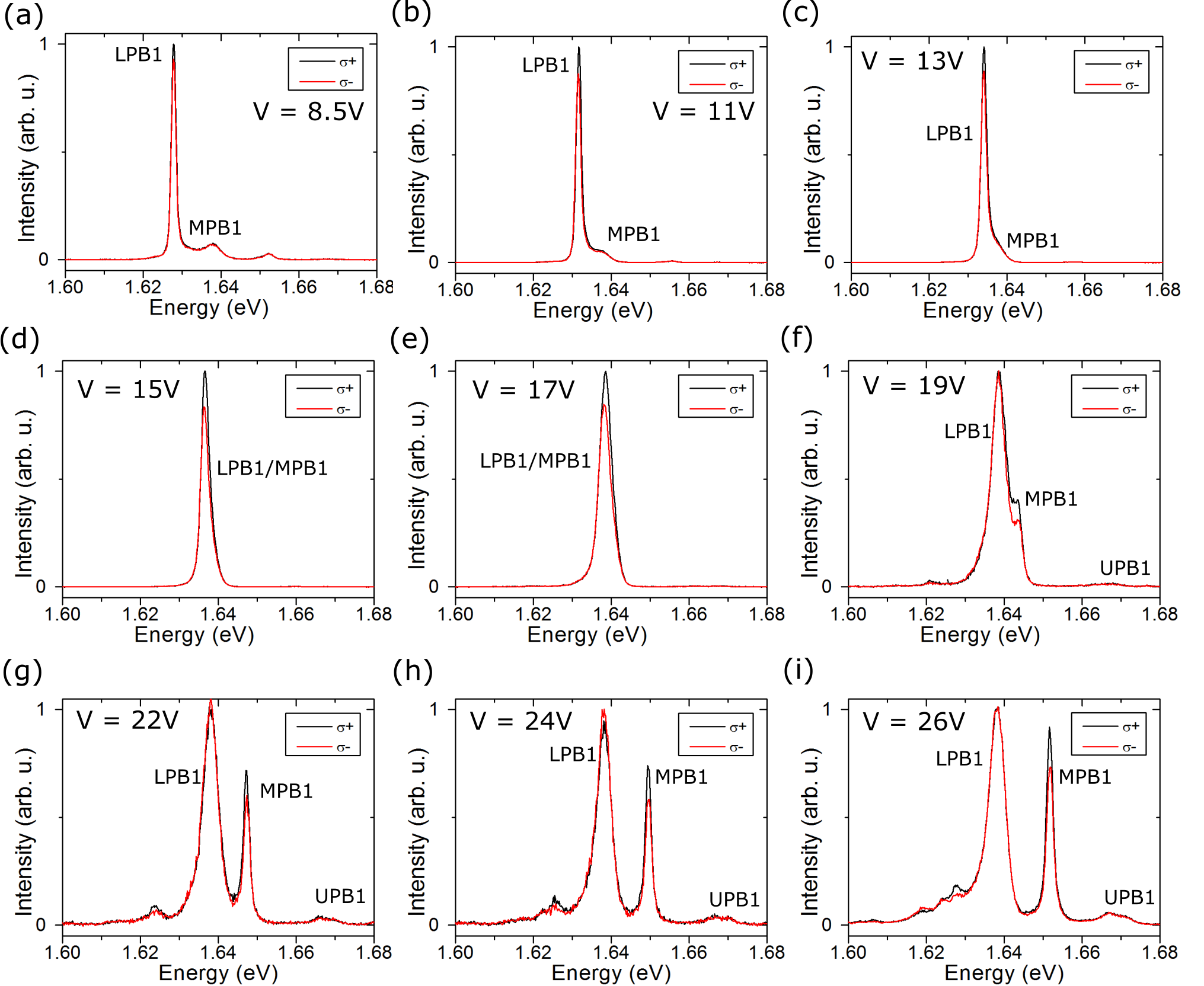}
\caption{\label{fig:fig2} Polarization resolved spectra when the cavity mode is tuned through the MoSe$_2$ trion resonance (z-piezo voltages from 8.5V to 26V). Piezo voltages correspond to slices from in Figure 2b of main text. (a) 8.5V (b) 11V (c) 13V (d) 15V (e) 17V (f) 19V (g) 22V (h) 24V (i) 26V.}
\end{figure}

\clearpage

\begin{figure}[h]
\includegraphics{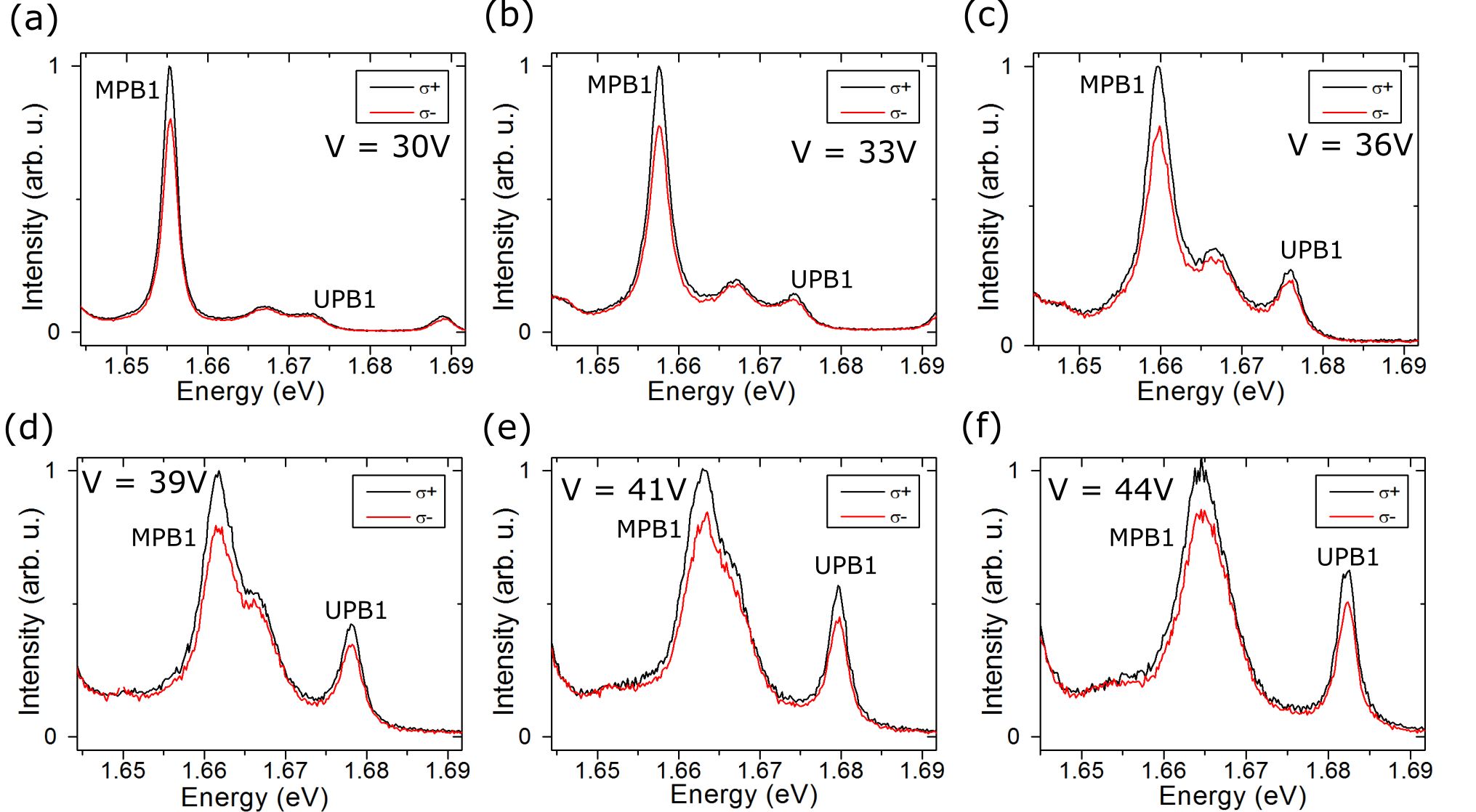}
\caption{\label{fig:fig3} Polarization resolved spectra when the cavity mode is tuned through the MoSe$_2$ neutral exciton resonance. Piezo voltages correspond to slices from in Fig. 2a of main text. (a) 30V (b) 33V (c) 36V (d) 39V (e) 41V (f) 44V. Polariton branches can clearly be resolved with a Rabi splitting of 15.2 meV which is significantly larger than the polariton linewidths. Varying degrees of retained valley polarisation are present for the polariton branches which strongly depend upon the exciton-photon detuning as discussed in the main text.}
\end{figure}

\clearpage

\section{Supplementary Note 3: Data Fitting Procedure}

\subsection{Estimate of Rabi splitting and Polarisation Degree}
Polariton branches are fitted with lorentzian functions in order to extract the peak positions to estimate the Rabi splitting. The Hamiltonian describing the coupling between the exciton, trion and photon is given by:

\begin{equation}
	H = \begin{pmatrix}
	E_X & 0 & V_X\\
	0 & E_T & V_T\\
	V_X & V_T & E_C \\
	\end{pmatrix}
	\label{eq:supp_hamiltonian}
\end{equation}
\noindent
where $E_X$, $E_T$ and $E_C$ are the exciton, trion and cavity energies respectively. $V_X$ and $V_T$ are the exciton-photon and trion-photon coupling strengths. 

The peak positions are obtained from a fit of multiple lorentzians to the various peaks using a non-linear least squares method. The peak positions are then simultaneously fitted with the analytical solution to the Eqn.~\ref{eq:supp_hamiltonian} where all parameters are shared in the fitting procedure. The cavity mode energy is assumed to follow the relationship $E_c = y_0 + aV + bV^2$ where the coefficients $a$ and $b$ take into account the linear and nonlinear change in mirror separation as a function of applied piezo voltage (V). The resultant fit is shown in Fig. S8 where the fitted Rabi splittings for trion-cavity and exciton-cavity couplings are $0.5 \pm 0.5$ meV and $15.2 \pm 0.1$ meV respectively. LPB and MPB peak positions are poorly reproduced around the trion resonance. A more accurate estimate of the trion-cavity coupling strength is given in Fig. S2 where at resonance the fitted LPB/MPB separation at zero trion-cavity detuning is around $1.3 \pm 0.1$ meV. The polarization degree in Fig. 2 of the main text is calculated from the peak-intensity extracted from the fits of the co- and cross-polarized data. In Fig. 2 of the main text some MPB data points at slight negative trion-cavity detuning have been discarded. These correspond to fits which which poorly reproduce the MPB intensity due due to the order of magnitude difference in intensity between LPB and MPB at around $-30$ meV.

\begin{figure}
\includegraphics{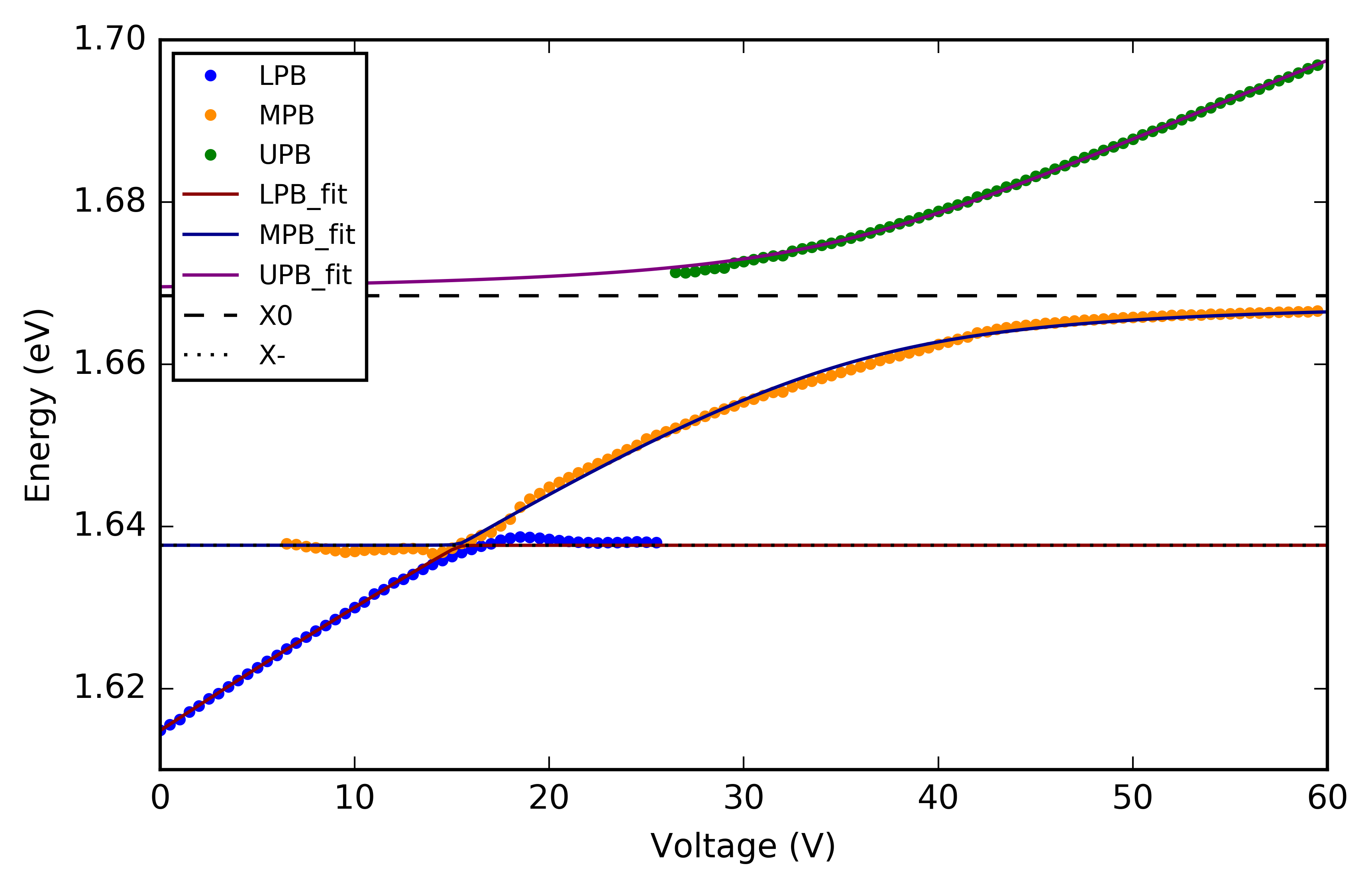}
\caption{Polariton peak energies fitted with 3-level coupled oscillator model.}
\end{figure}

\clearpage

\begin{figure}
\includegraphics{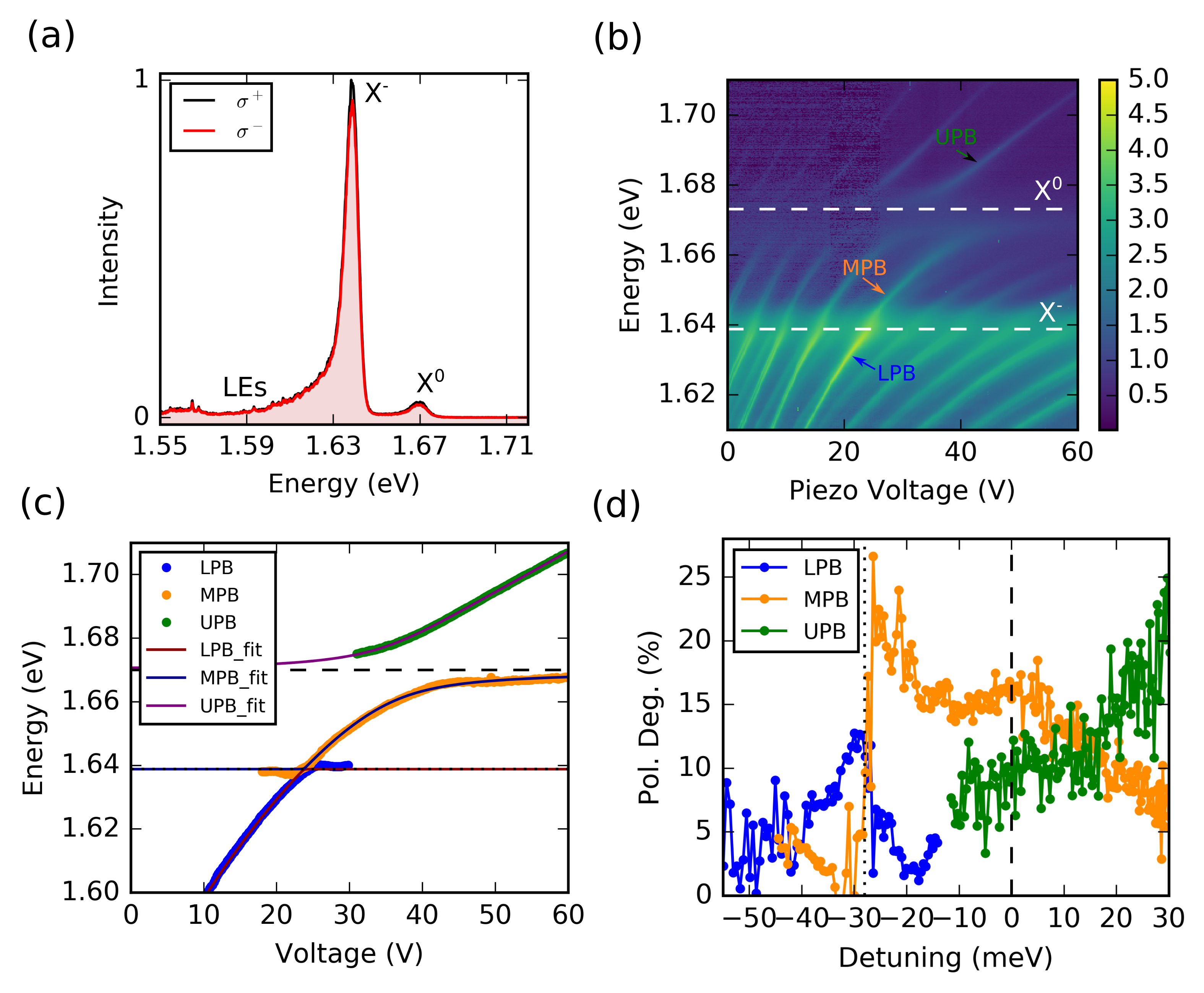}
\caption{Data for second sample consisting of hBN/MoSe2 heterostructure. (a) PL spectra of MoSe$_2$ under $\sigma^+$ excitation and co- (black) and cross- (black) polarized detection. The polarization degree is 8\% and 3\% for the exciton and trion respecively.  (b) Anticrossing between the ground state $LG_{00}$ mode and the exciton energy. (c) Fitted peak positions giving neutral exciton Rabi splitting of $18.0 \pm 0.1$ meV. Fitting the LPB/MPB positions around the trion resonance is poorly reproduced. A more accurate estimate of the trion-cavity coupling is given through similar fitting as for sample 1 in Fig. S2 which gives a separation between LPB and MPB at resonance of $1.5 \pm 0.2$ meV. (d) Polarisation degree of the polariton branches as a function detuning.}
\end{figure}

\clearpage

\begin{figure}
\includegraphics{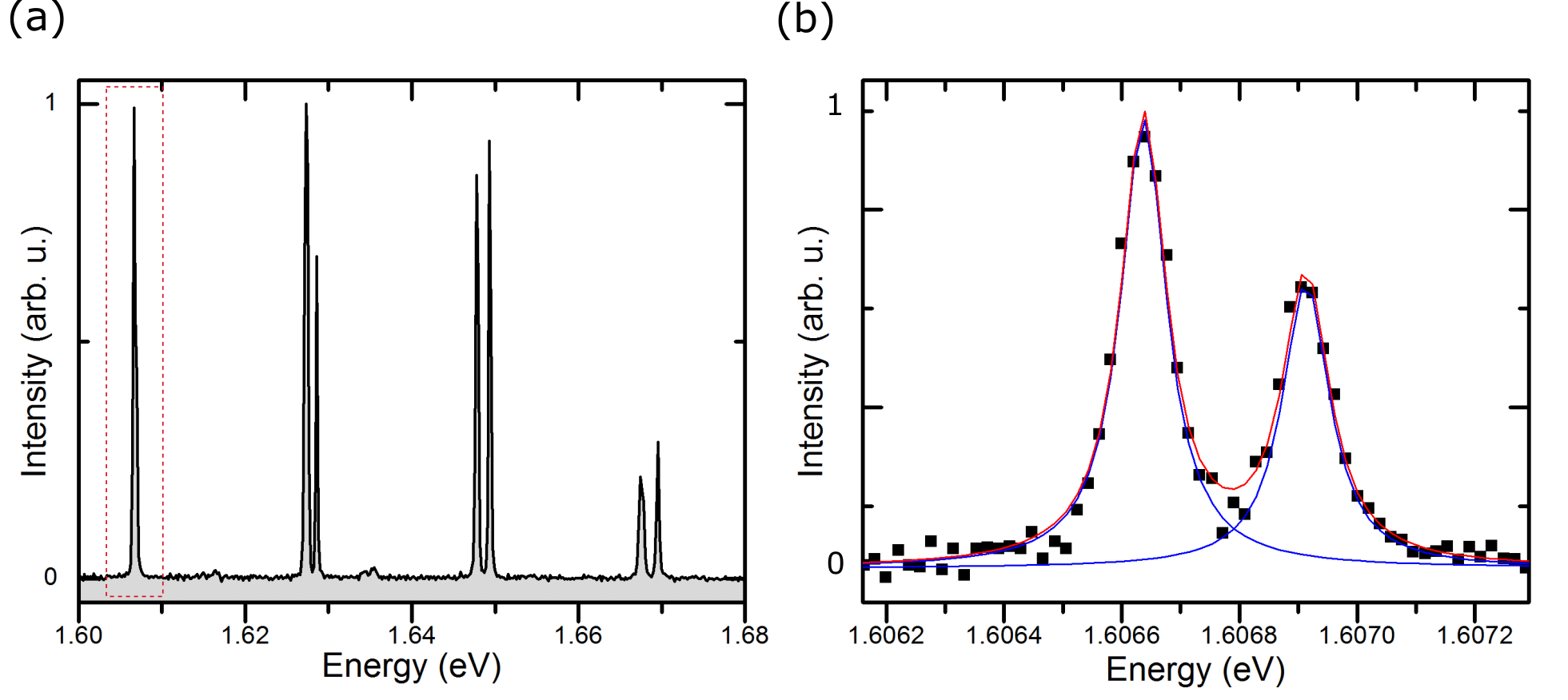}
\caption{\label{fig:fig4} (a) Cavity mode structure showing ground state (red box) and higher order transverse modes, which arise due to the harmonic like photonic confinement potential, in the weak coupling regime. (b) Spectrum of the TE-TM split ground state mode. The TE-TM splitting is 270 \micro eV.}
\end{figure}

\clearpage

\begin{figure}
\includegraphics{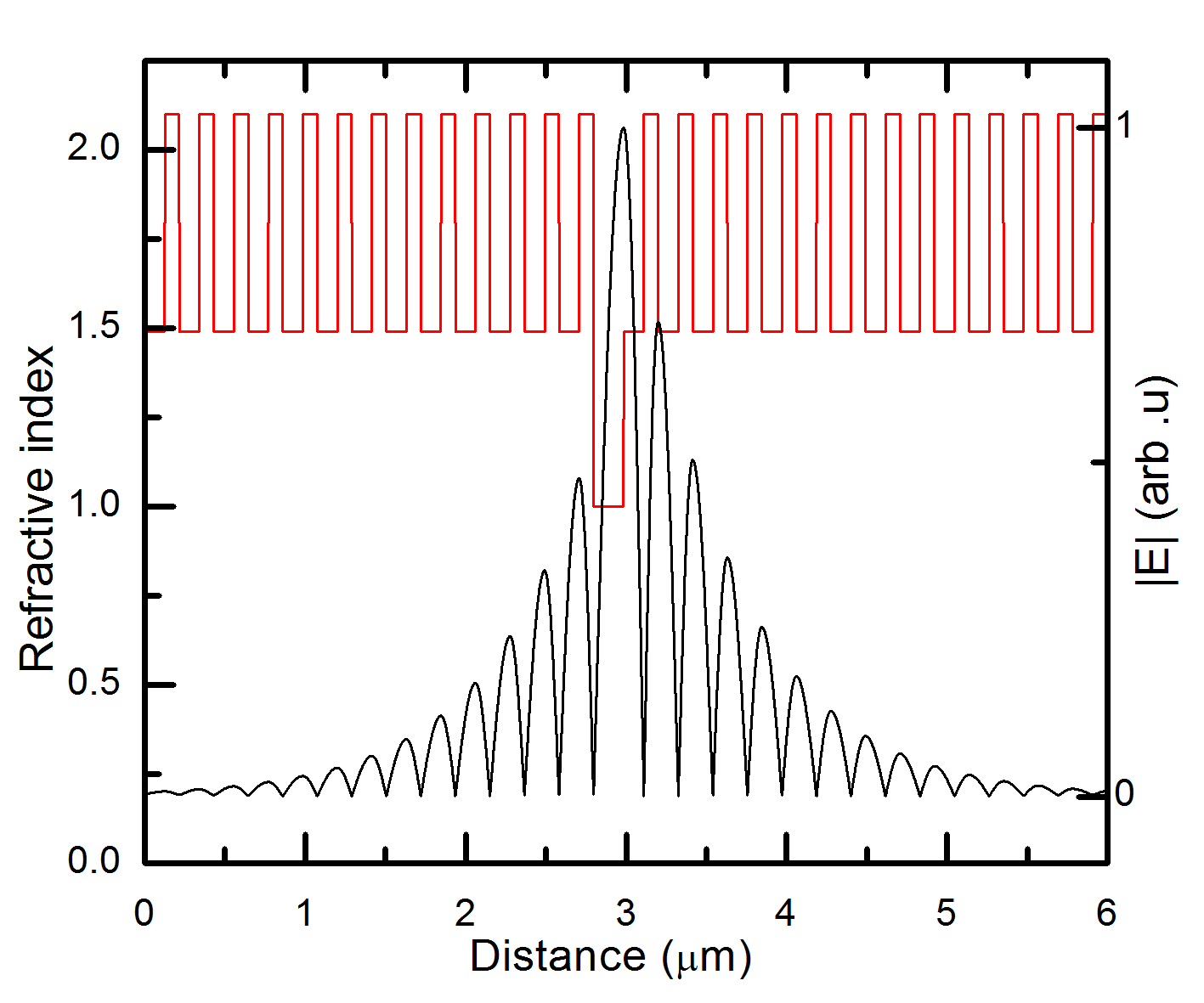}
\caption{\label{fig:fig1} Electric field distribution of microcavity calculated using the transfer matrix method. The formed tunable microcavity consists of two dielectric DBRs seperated by a $q\lambda/4n$ gap, where q is the longitudinal mode index. The top concave DBR terminates with high index material giving a node at the surface while the bottom planar DBR terminates with low index giving an antinode. The TMD monolayer is placed at the surface of the bottom DBR at an electric-field antinode. The DBRs used in the main manuscript were 10 paired SiO$_2$/Nb$_2$O$_5$.}
\end{figure}

\clearpage

\begin{figure}
\includegraphics[scale=0.95]{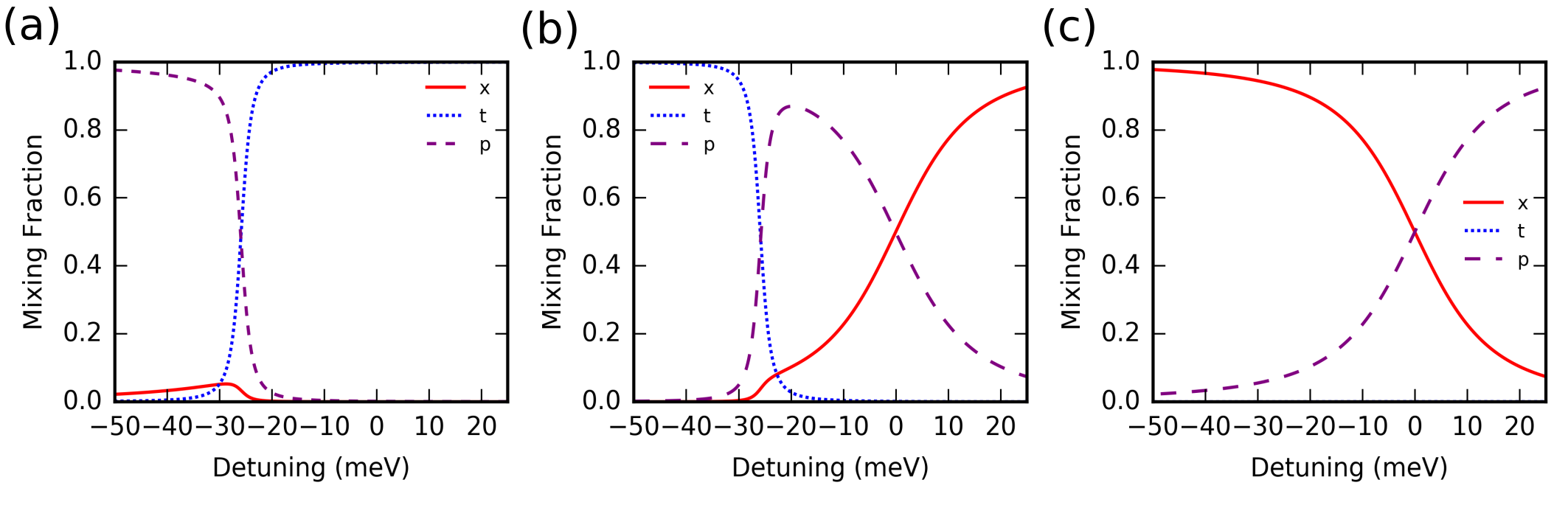}
\caption{\label{fig:fig5} Hopfield coefficients for the three polariton branches (a) LPB (b) MPB and (c) UPB used in the calculations of polarization degree as a function of detuning shown in the main text.}
\end{figure}

\end{document}